\documentclass[pdflatex,sn-nature]{sn-jnl}% Style for submissions to Nature Portfolio journals
\usepackage{graphicx}
\usepackage[justification=justified,singlelinecheck=false]{caption}
\usepackage{float}
\usepackage{subcaption} 
\usepackage{longtable}
\usepackage{multirow}
\usepackage{amsmath,amssymb,amsfonts}%
\usepackage{amsthm}%
\usepackage{mathrsfs}%
\usepackage[title]{appendix}%
\usepackage{xcolor}%
\usepackage{textcomp}%
\usepackage{manyfoot}%
\usepackage{booktabs}%
\usepackage{algorithm}%
\usepackage{algorithmicx}%
\usepackage{algpseudocode}%
\usepackage{listings}%
\theoremstyle{thmstyleone}%
\theoremstyle{thmstyletwo}%

\theoremstyle{thmstylethree}%

\begin{document}

\title[Article Title]{Mass-radius relation, moment of inertia, and tidal love numbers of anisotropic neutron stars in $f\left(R,T\right)$ gravity}

%%=============================================================%%
%% GivenName	-> \fnm{Joergen W.}
%% Particle	-> \spfx{van der} -> surname prefix
%% FamilyName	-> \sur{Ploeg}
%% Suffix	-> \sfx{IV}
%% \author*[1,2]{\fnm{Joergen W.} \spfx{van der} \sur{Ploeg} 
%%  \sfx{IV}}\email{iauthor@gmail.com}
%%=============================================================%%

\author[1,2]{\fnm{Yusmantoro Yusmantoro}}\email{yusmantoroyusman@gmail.com}

\author[1,2]{\fnm{Freddy} \sur{Permana} \sur{Zen}}\email{fpzen@fi.itb.ac.id}
%\equalcont{These authors contributed equally to this work.}
  
\author[1,2]{\fnm{Muhammad} \sur{Lawrence} \sur{Pattersons}}\email{m.pattersons@proton.me}
%\equalcont{These authors contributed equally to this work.}

\affil[1]{\orgdiv{Theoretical High Energy Physics Group}, \orgname{Department of Physics, FMIPA, Institut Teknologi Bandung}, \orgaddress{\street{Jl. Ganesha 10}, \city{Bandung}, \country{Indonesia}}}

\affil[2]{\orgdiv{Indonesia Center for Theoretical and Mathematical Physics (ICTMP)}, \orgname{Institut Teknologi Bandung}, \orgaddress{\street{Jl. Ganesha 10}, \city{Bandung}, \postcode{40132}, \country{Indonesia}}}

%\affil[3]{\orgdiv{Research Center for Quantum Physics}, \orgname{National Research and Innovation Agency (BRIN)}, \orgaddress{\city{South Tangerang}, \postcode{15314}, \country{Indonesia}}}

%%==================================%%
%% Sample for unstructured abstract %%
%%==================================%%

\abstract{The mass-radius relation, moment of inertia, and tidal love numbers of anisotropic neutron stars (NSs) have been investigated in $f(R,T)$ gravity by imposing two equations of state (EoS). We use the simplest form $f(R,T)=R+2\beta T$ model and adopt the anisotropy approach called Horvat model. To examine the viability of our calculations, we utilize the constraints from GW170817 and GW190814 observations. Moreover, we consider three values of $\beta$, i.e. $\beta=0$, $\beta=-0.01$, $\beta=-0.02$ and four anisotropy parameters $\alpha$, i.e. $\alpha=-0.12$, $\alpha=-0.06$, $\alpha=0.06$, $\alpha=0.12$. Our findings suggest that all physical quantities depend on both parameters $\alpha$ and $\beta$. Nevertheless, the impact of $\alpha$ is much more significant than $\beta$. The calculation of masses satisfy each used constraints for specific values of $\alpha$ and $\beta$. In the case of the moment of inertia, the results are compatible with the constraint obtained from radio observation of heavy pulsar. On the other hand, the tidal deformability of the NSs composed of one EoS satisfy the  GW170817 constraint while the NSs composed of the other one EoS are too small. These small numbers can be interpreted as the property of secondary object observed in GW190814. As a result, our theoretical investigation of NSs constructed with two EoS can be NSs candidates for GW170817 and GW190814, respectively.}


%%\pacs[JEL Classification]{D8, H51}

%%\pacs[MSC Classification]{35A01, 65L10, 65L12, 65L20, 65L70}

\maketitle

\section{Introduction}\label{sec1}
Neutron stars (NSs) are one of the most interesting astrophysical objects to study because their physical properties are more complex to characterize through observation than other objects such as black holes and strange stars\citep{lattimer2007neutron}. For instance, gravitational waves (GWs) detection of binary NSs mergers conducted by advanced detectors like LIGO, Virgo, and KAGRA has opened possibility to investigate the stellar nucelar matter by putting constraints on its EoS\citep{anderson2008magnetized,bauswein2012measuring,rodriguez2014basic,berry2015parameter,burns2020neutron,kagra2020prospects}. Consequently, other quantities like moment of inertia, tidal deformability can be examined. Unlike black holes which have to obey the no-hair theorem, physics laws allow NSs to have more complex structures and configurations. As a result, specific attemps have been performed by Neutron Star Interior Composition Exploler (NICER) to probe the intrinsic features of them including mass-radius relation, compactness, and spin frequency. Accordingly, these results give rise to the rapid development of theoretical studies in astrophysics \citep{gendreau2012neutron,miller2019psr,yamamoto2023quark,brandes2025implications}.
  
The observation of GW190814 by the LIGO and Virgo collaborations in 2019 confirms that this signal is emitted by a merger of two compact objects having quite different masses. The first stellar remnant of mass between $\left(22.2-24.3\right)M_{\odot}$ was a black hole while the second one with mass of $\left(2.50-2.67\right)M_{\odot}$ is still unidentified \citep{harry2020gw190814}. Some physicists belive that the second object was a small black hole \citep{clesse2022gw190425,lyu2023revisiting} while others think it was a NS \citep{huang2020possibility,dexheimer2021gw190814}. If the object wass a black hole, its mass was too small and it would be the lightest black hole ever detected. However, if it was a NS, it was the most massive NS ever observed. Therefore, the identification of this object remains a mystery.

Some researches are performed to construct a theoretical model of mass of NSs in General Relativity as well as its extension \citep{astashenok2015extreme,astashenok2020extended}. In Ref. \citep{pattersons2021mass}, the total mass of anisotropic NSs is calculated using Hartle-Thorne formalism. They show that anisotropy and rotation significantly increase the mass of NSs up to more than $2.25\hspace{1mm}M_{\odot}$. 
Subsequently, this work is extended and developed in Ref.
\citep{beltracchi2024slowly} by considering comprehensive study of integral and surface features of Bowers-Liang anisotropy model.
The same formalism is also conducted in Rastall gravity to calculate the rotational mass of anisotropic NSs \citep{pattersons2025rotational}. In their work, the Rastall and anisotropy parameters successfully enhance the mass of NSs up to $2.67\hspace{1mm}M_{\odot}$ which corresponds to NSs in GW190814 observation.

Another measurable quantity comes from GWs observation of the merging NSs is the tidal love numbers (TLNs). These variables specify the induced multipole moments or deformation of NSs when they are immersed in an external tidal field. Information about TLNs is carried by shifting phase of GWs of pre-merger inspiral stage\citep{hinderer2008tidal,flanagan2008constraining,binnington2009relativistic}. Due to direct relation between TLNs and internal structure of NSs, they are believed to exclude a considerable number of equations of state. In addition, TLNs also capable of evaluating a theory of gravity in a region of intense gravitational field \citep{cardoso2017testing,franzin2017testing}. Thus, exploring TLNs of NSs will provide a significant and valuable impacts on astrophysical investigation.

The TLNs were firstly formulated in Newtonian framework of the Earth with respect to disturbances in the form of tidal forces \citep{love1911some}. In that formulation, two parameters $h$ and $l$ were introduced to characterize vertical and horizontal deformation of a spherically symmetric object. Moreover, the third variable $k$ is defined to specify the distortion of the gravitational potential \citep{love1909yielding}. Equivalently, $k$ relates the disturbance and the additional gravitational potential of the Earth which emerges as a physical response to the external forces. A concrete example commonly observed is the tides of the sea due to the gravitational force of the moon. The larger $k$ indicates a greater response to external fields, which means that the object is more flexible.Conversely, a smaller $k$ means that the object is more robust and more difficult to be deformed \citep{love1909yielding}.

In the case of stellar objects like NSs or black holes, the TLNs have to be formulated relativistically because the gravitational field around them are extremely strong. For this reason, Newtonian theory does not hold accurately and should be replaced by  GR. Relativistic formulation of TLNs for NSs and black holes has been established in \citep{hinderer2008tidal,flanagan2008constraining,binnington2009relativistic}. It has been shown that black holes do not undergo deformation due to external fields and this is what distinguishes them from NSs. The zero value of TLNs is suspected because black holes are enclosed by event horizon which is an inelastic surface that absorbs the tidal deformation and prevents the emergence of static multipoles \citep{hinderer2008tidal}.

In relativistic formalism, metric tensor acts like gravitational potential in Newtonian theory where it encodes complete information about the dynamics of spacetime. Hence, TLNs in GR are investigated using metric perturbation which results in linearized Einstein gravity. From the solution of such equation, we will obtain the perturbation function $H(r)$ that is interpreted as the wave carrying tidal response of NSs. Variabel $H$ consist of two parts namely growing and decaying terms that stand for source of external field and the response of NSs, respectively. Then, TLNs is defined as the ratio between the decaying and the growing components of $H(r)$.

According to Regge-Wheeler formalism, metric perturbation of a spherically symmetric NSs can be decomposed into two components i.e. even (polar) and  odd (axial) parities  \citep{zerilli1970effective,regge1957stability}. Hence, the corresponding results of each calculations are called  polar (electric) $k^{\mathrm{polar}}$ and axial (magnetic) $k^{\mathrm{axial}}$ TLNs, respectively. Polar TLNs quantify how geometric of a stellar object is deformed. In the case of merging NSs, such quantities are symmetric relative to the axis connecting both stars. On, the other hand, axial TLNs measure the change of multipole current inside the NSs and are not related to the shape of them.  

The TLNs of NSs in GR and their possibility to be probed by Advanced LIGO observation has been explored comprehensively \citep{hinderer2010tidal,damour2009relativistic}. In the same framework, TLNs of stars composed of exotic matters namely axion, boson, and proca stars are investigated and the results suggest that electric TLNs are positive, while magnetic TLNs are negative \citep{cardoso2017testing,chen2024tidal,herdeiro2020tidal}. TLNs of a compact star toy-model composed of a linear equation of state have also been calculated by considering anisotropy \citep{parida2023toy}. The impacts of elastic crust  and dissipative effects on the deformation of NSs and have been studied in Ref. \citep{biswas2019role,ripley2023probing}.
Universal relation, specifically I-Love-C and I-Love-Q relations which connect moment of inertia, TLNs, and compactness or tidal deformability of NSs has been revealed in Einstein theory \citep{das2022love,yeung2021love}. Based on those relations, there are three corresponding properties of NSs that do not depend on the equation of state. If one quantity is determined, the two remaining two can be calculated automatically.

Based on the above description, GR appears to be very good at determining the TLNs and other properties of stellar objects like NSs, pulsar, and exotic stars. Nevertheless, GR is not capable of solving major problem in cosmology such as dark matter and dark energy. Hence, applying the extension of Einstein's theory in cosmological scale is very crucial \citep{de2012black}. Therefore, it is essential to examined TLNs of NSs within modified gravity (MG) so that astrophysics and cosmology can be studied in the single framework. Additionally, MG provides new degrees of freedom appearing as an advantage of it over GR and these are used to characterize the dynamics of the universe. Thus, analyzing the impact of them on features of astrophysical bodies like NSs is fascinating \citep{saito2015modified,yazadjiev2014non,staykov2014slowly,yazadjiev2015rapidly,doneva2015iq,capozziello2016mass,yazadjiev2017oscillation,doneva2018differentially,olmo2020stellar,feola2020mass,astashenok2021novel,odintsov2021neutron,odintsov2022neutron,astashenok2023chandrasekhar,odintsov2023inflationary,oikonomou2023rp,oikonomou2023static}.

Calculation of magnetic and electric TLNs of NSs filled with polytropic matter in Rastall gravity are smaller than those in GR. It implies that Rastall parameter make NSs more difficult to be deformed \citep{meng2021tidal}. Polar and axial TLNs of NSs in Starobinsky $f(R)$ model have also been examined and it has been shown that polar TLNs are slightly different compared to that of GR while axial TLNs are much more larger \citep{yazadjiev2018tidal}. In contrast, magnetic TLNs of NSs in unimodular gravity are smaller than GR while electric-type ones exceed those in GR \citep{yang2022tidal}. Tidal effects represented by tidal deformability consisting of scalar, tensor, and mixed sector on GWs emitted by binary system of NSs have been shown in scalar-tensor theories  \citep{creci2025tidal}.

The study of NSs with a realistic EoS in MG is essential. Moreover, the pressure of nuclear matter inside them is generally not isotropic. Accordingly, considering the effect anitropy on the features of NSs will make the theoretical investigation becomes more viable and convincing. In this research, we investigate the properties of anisotropic NSs in $f(R,T)$ gravity with two equations of state namely quantum hadrodynamics (QHD) \citep{lopes2020role} and BPS+$\beta$ EoS\citep{whittenbury2015hadrons,whittenbury2014quark,whittenbury2016hybrid}. In the QHD model, baryons are fundamental degrees of freedom where their interractions are considered via meson exchange. Meanwhile, $\mathrm{BPS}+\beta$ framework contains low and high density EoS called BPS (crust part) \citep{baym1971ground} and $\beta-\mathrm{stable}$ (core part) derived from a Hartree-Fock of the Quark-Meson Coupling model.

Initially, $f(R,T)$ model was proposed by Harko et al. \citep{harko2011f} as  a generalization of $f(R)$ gravity. They aim to involve the coupling between geometry and matter represented by Ricci scalar $R$ and the trace of energy-momentum tensor $T$, respectively. This model has also been applied to investigate the gravitational signature of black holes using nonlinear electrodynamics \citep{araujo2025gravitational}. We consider the simplest $f(R,T)$ theory i.e. $f(R,T)=R+2\beta T$ where $\beta$ is a free parameter. Such $f(R,T)$ model has been studied to examine equilibrium configuration of compact stars \citep{das2016compact,deb2018strange,biswas2021anisotropic,pappas2022extended}.
The moment of inertia and TLNs of NSs has also been investigated in Ref.\citep{pretel2022moment,murshid2024neutron}.  
Additionally, we use anisotropic model proposed by Horvat et al. \citep{horvat2010radial} which is given by $\sigma=2\alpha m p_{r}/r$.  The free parameter $\alpha$ denotes the anisotropic strength. 

In this work, we calculate the static mass of NSs and use two constraints from GW170817 $\left(1.36-2.26\right)M_{\odot}$ and GW190814 $\left(2.50-2.67\right)M_{\odot}$\citep{abbott2017gw170817,harry2020gw190814}.
We also consider additional constraints from GW170817 namely the excluded areas in $M$ vs $r$ plane for NSs and the allowed radii for NSs with mass $1.4\hspace{1mm}M_{\odot}$ and $2.0\hspace{1mm}M_{\odot}$ \citep{bauswein2017neutron,altiparmak2022sound}.
Then, we examine the moment of inertia by taking slowly rotation limit of NSs' metric. To examine viability, we employ the constraint from Landry et al.\citep{landry2020nonparametric}. Moreover, we calculate the TLNs and tidal deformability of NSs using metric perturbation. We provide constraints of the tidal deformability of the binary NSs merger GW170817 \citep{abbott2017gw170817,abbott2018gw170817}. The tidal deformability of a NS with mass of $M=1.4\hspace{1mm}M_{\odot}$ is constrained to be $580<\Lambda_{1.4}<800$. The numerical computations in this research are conducted using the CompactObject package from Huang et al\citep{huang2024compactobject}.

One of the novel results of our work compared to the previous one \citep{pattersons2025rotational} is that we obtain the high mass value of NSs with QHD EoS without considering rotation effect. This corresponds to secondary object of GW190814 and therefore our result support the prediction that such object is a NS. Our result is also more physical due to the fact that the mass obtained in GWs observaton is commonly a static mass not a rotational mass. Furthermore, we enhance our argument from TLNs and tidal deformability calculations which are quite small $\left(\Lambda<200\right)$ compared to that of common NSs. That is why the tidal deformability of secondary object is undetected.
Another novel finding of this study is that the NSs with $\mathrm{BPS}+\beta$ EoS depicts the mass of NSs in GW170817 observation \citep{abbott2017gw170817}. Additionally, the tidal deformability curve at a mass of $1.4\hspace{1mm}M_{\odot}$ lies between $750-800$ for specific $\alpha$. The plots of moment inertia for both Eos also satisfies the constraint from Landry et al.\citep{landry2020nonparametric}.

\section{Field equations in $f\left(R,T\right)$ gravity}\label{sec2}
The action of modified $f(R,T)$ gravity takes the following form \citep{harko2011f}
\begin{equation}\label{eq:01}
	 S=\frac{1}{16\pi}\int\sqrt{-g} f(R,T) \,d^{4}x +\int\sqrt{-g} \mathcal{L}_m\,d^{4}x,
\end{equation}
where $f(R,T)$ in the first term of right hand side in equation \ref{eq:01} is the combination of Ricci scalar $R$ and the trace of energy-momentum tensor $T$. While the second term $\mathcal{L}_m$ represents the lagrangian of matter. To obtain the field equations of $f(R,T)$ gravity, we take the variation of above action with respect to metric tensor $g^{\mu\nu}$ which yields equation \ref{eq:02}
\begin{equation}\label{eq:02}
	\begin{split}
		\frac{\partial f(R,T)}{\partial R}R_{\mu\nu}+&\left(g_{\mu\nu}\Box-\nabla_{\mu}\nabla_{\nu}\right)\frac{\partial f(R,T)}{\partial R}+\frac{\partial f(R,T)}{\partial T}\left(T_{\mu\nu}+\Theta_{\mu\nu}\right)\\&- \frac{1}{2}f(R,T)g_{\mu\nu}
		=8\pi T_{\mu\nu}.
	\end{split}
\end{equation}
$\Box=\nabla^{\mu}\nabla_{\nu}$ is the d'Alembertian operator and $T_{\mu\nu}$ is the energy-momentum tensor which is defined as 
\begin{equation}\label{eq:03}
	T_{\mu\nu}=-\frac{2}{\sqrt{-g}}\frac{\delta\left(\sqrt{-g}\mathcal{L}_m\right)}{\delta g^{\mu\nu}}.
\end{equation}
Then, tensor $\Theta_{\mu\nu}$ is described by
\begin{equation}\label{eq:04}
	\Theta_{\mu\nu}=g^{\alpha\beta}\frac{\delta T_{\alpha\beta}}{\delta g^{\mu\nu}}.
\end{equation}
Taking covariant derivative of equation \ref{eq:02} yields
\begin{equation}\label{eq:05}
	\nabla^{\mu}T_{\mu\nu}=\frac{\frac{\partial f(R,T)}{\partial T}}{8\pi-\frac{\partial f(R,T)}{\partial T}}\left[
	\left(T_{\mu\nu}+\Theta_{\mu\nu}\right)\nabla^{\mu}\ln\left(\frac{\partial f(R,T)}{\partial T}\right)+\nabla^{\mu}\Theta_{\mu\nu}-\frac{1}{2}g_{\mu\nu}\nabla^{\mu}T\right]
\end{equation}
In this paper, we take specific form of $f(R,T)$ as $f(R,T)=R+2\beta T$. This is a viable model that has been studied extensively in cosmology and astrophysics \citep{harko2011f}. Consequently, equations \ref{eq:02} and \ref{eq:05} transform to
\begin{equation}\label{eq:06}
	\begin{split}
		G_{\mu\nu}&=8\pi T_{\mu\nu}+\beta\left(Tg_{\mu\nu}-2T_{\mu\nu}-2\Theta_{\mu\nu}\right)\\
		R_{\mu\nu}-\frac{1}{2}g_{\mu\nu}R&=8\pi T_{\mu\nu}+\beta\left(Tg_{\mu\nu}-2T_{\mu\nu}-2\Theta_{\mu\nu}\right)
	\end{split}
\end{equation}
\begin{equation}\label{eq:07}
	\nabla^{\mu}T_{\mu\nu}=\frac{2\beta}{8\pi-2\beta}\left[\nabla^{\mu}\Theta_{\mu\nu}-\frac{1}{2}g_{\mu\nu}\nabla^{\mu}T\right].
\end{equation}
Moreover, we assume that the Lagrangian of matter is expressed $\mathcal{L}_{m}=\left(p_r+2p_t\right)/3$. $p_t$ is the tangential pressure and $p_r$ is the radial pressure. Consequently, tensor $\Theta_{\mu\nu}$, its trace $\Theta$, and the Ricci scalar $R$ can be written as
\begin{equation}\label{eq:08}
	\begin{split}
		\Theta_{\mu\nu}&=-2T_{\mu\nu}+\left(\frac{p_{r}+2p_{t}}{3}\right)g_{\mu\nu}\\
		\Theta&=-2T+\frac{4}{3}p_{r}+\frac{8}{3}p_{t}\\
		R&=-8\pi T-2\beta\left(3T-\frac{4}{3}p_r-\frac{8}{3}p_t\right).
	\end{split}		
\end{equation}

\section{TOV equations in $f(R,T)$ gravity}\label{sec3}
In this work, we take into account anisotropic property of pressure inside the NS meaning that radial pressure $p_r$ and tangential pressure $p_r$ are unequal. Therefore, the energy-momentum reads
\begin{equation}\label{eq:09}
	T_{\mu\nu}=\left(\rho+p_t\right)u_{\mu}u_{\nu}+p_{t}g_{\mu\nu}-\sigma k_{\mu}k_{\nu},
\end{equation}
where $u^{\mu}=e^{-\psi}\left(1,0,0,0\right)$ is the four-velocity and $k^{\mu}=e^{-\lambda}(0,1,0,0)$ is the unit radial four-vector. These vectors satisfy $u_{\mu}u^{\mu}=-1$ and $k_{\mu}k^{\mu}=1$, respectively. 
The energy density is denoted by $\rho$ and the last term $\sigma=p_t-p_r$ stands for the anisotropy factor. Consequently, the trace of $T_{\mu\nu}$ is given by $T=-\rho+3p_r+2\sigma$.
 
Furthermore, we consider a static and spherically symmetric metric of neutron star. The line element in Schwarzschild coordinate $(t,r,\theta,\phi)$ can be written as
\begin{equation}\label{eq:10}
	ds^2=-e^{2\psi}dt^2+e^{2\lambda}dr^2+r^2d\theta^2+r^2\sin^2{\theta}d\phi^2.
\end{equation}
We assume that metric potentials $\psi$ and $\lambda$ depend only on radial coordinate $r$. Hence the mass of neutron star is the function or $r$ only as well and its relation with $\lambda$ can be expressed as
\begin{equation}\label{eq:11}
	e^{2\lambda}=\left(1-\frac{2m}{r}\right)^{-1}.
\end{equation}

Considering $tt$-component of the field equation and putting equation \ref{eq:10} to it, we have 
\begin{equation}\label{eq:12}
	\frac{dm}{dr}=4\pi r^2\rho+\frac{\beta r^2}{2}\left(3\rho-p_r-\frac{2}{3}\sigma\right).
\end{equation}
Meanwhile, the differential equation of metric potential $\psi$ can be derived by analyzing $rr$-component of equation \ref{eq:06}
\begin{equation}\label{eq:13}
	\frac{d\psi}{dr}=\left[\frac{m}{r^2}+4\pi rp_r+\frac{\beta r}{2}\left(-\rho+3p_r+\frac{2}{3}\sigma\right)\right]\times\left(1-\frac{2m}{r}\right)^{-1}.
\end{equation}
The equation of radial pressure is obtained by examining radial component of equation \ref{eq:07}
\begin{equation}\label{eq:14}
	\frac{dp_r}{dr}=-\left(\rho+p_r\right)\psi'+\frac{2}{3}\sigma+\frac{\beta}{8\pi+2\beta}\frac{d}{dr}\left(\rho-p_r-\frac{2}{3}\sigma\right).
\end{equation}
Substituting equation \ref{eq:13} to \ref{eq:14} yields the hydrostatic equilibrium equation of the star
\begin{equation}\label{eq:15}
	\begin{split}
		\frac{dp_r}{dr}=&-\frac{\rho+p_r}{1+\frac{\beta}{8\pi+2\beta}}\left[\frac{m}{r^2}+4\pi rp_r+\frac{\beta r}{2}\left(-\rho+3p_r+\frac{2}{3}\sigma\right)\right]\times\left(1-\frac{2m}{r}\right)^{-1}\\
		&+\frac{\frac{\beta}{8\pi+2\beta}}{1+\frac{\beta}{8\pi+2\beta}}
		+\frac{2}{1+\frac{\beta}{8\pi+2\beta}}\left(\frac{\sigma}{r}-\frac{\frac{\beta}{8\pi+2\beta}}{3}\frac{d\sigma}{dr}\right).
	\end{split}
\end{equation}
Equations \ref{eq:12} and \ref{eq:15} are TOV equations in $f(R,T)$ gravity.

There is no matter outside the region of the star and we can write $T=p_r=p_t=\rho=0$ yielding $T_{\mu\nu}=0$. Hence, based on equations \ref{eq:06} and \ref{eq:12}, we have $G_{\mu\nu}=0$ and $dm/dr=0$. These conditions imply that the geometry of spacetime outside the star can be described using Schwarzschild vacuum solution.
The interior and exterior spacetimes are smoothly connected at $r=r_{\mathrm{surface}}$. Therefore, TOV equations have to satisfy the following boundary conditions
\begin{equation}\label{eq:015}
	\begin{split}
		m(0)&=0,\\
		m\left(r_{\mathrm{surface}}\right)&=M,\\
		e^{2\psi\left(r_{\mathrm{surface}}\right)}&=1-\frac{2M}{r_{\mathrm{surface}}},
	\end{split}	
\end{equation} 
where $M$ is the total mass of the star. Comparing our model with the previous works are totally different in term of describing spacetime outside the stars \citep{oikonomou2021uniqueness,oikonomou2021universal}.  In those references, the scalar field does exist beyond the surface of the star. Therefore, the metric beyond the surface is not directly a Schwarzschild due to the nonvanishing energy-momentum tensor $T_{\mu\nu}$. On the other hand, the $f(R,T)$ model considered in this paper is the same as GR outside the star.

\section{Moment of inertia}\label{sec4}
To calculate the moment of inertia of NSs, we use slowly rotating approximation. It implies that the static metric is extended to have rotation variable $\omega(r,\theta)$ denoting frame dragging rate of the spacetime
\begin{equation}\label{eq:16}
	ds^2=-e^{2\psi}dt^2+e^{2\lambda}dr^2+r^2d\theta^2+r^2\sin^2{\theta}d\phi^2-2\omega(r,\theta)r^2\sin^2{\theta} dtd\phi.
\end{equation}
Due to this approximation, the four-velocity of fluid changes into $u^{\mu}=u^{t}\left(1,0,0,\Omega\right)$, where $\Omega$ is the rotational velocity of the fluid with respect to the distant observer. Hence, $t\phi$-component of the mixed energy-momentum tensor can be expressed as
\begin{equation}\label{eq:17}
	T^{t}_{\phi}=\left(\rho+p_{t}\right)e^{-2\psi}r^{2}\sin^{2}\theta\left(\Omega-\omega\right).
\end{equation}

Contracting equation \ref{eq:06}, we get the following mixed tensor. 
\begin{equation}\label{eq:18}
	\begin{split}
		g^{\alpha\mu}G_{\mu\nu}&=8\pi g^{\alpha\mu}T_{\mu\nu}+\beta Tg^{\alpha\mu}g_{\mu\nu}-2\beta\left(g^{\alpha\mu}T_{\mu\nu}+g^{\alpha\mu}\Theta_{\mu\nu}\right)\\
		G^{\alpha}_{\nu}&=\left(8\pi+2\beta\right)T^{\alpha}_{\nu}+\beta\left(-\rho+\frac{1}{3} p_r+\frac{2}{3} p_t\right)\delta^{\alpha}_{\nu}.
	\end{split}
\end{equation}
$t\phi$-component of equation \ref{eq:18} can be written as
\begin{equation}\label{eq:19}
	\begin{split}
		R^{t}_{\phi}&=\left(8\pi+2\beta\right)T^{t}_{\phi}\\
		-\frac{e^{-(\lambda+\psi)}}{2r^2\sin\theta}\partial_{r}\left[e^{-(\lambda+\psi)}r^{4}\sin^{3}\theta\frac{\partial\omega}{\partial r}\right]&-\frac{e^{-(\lambda+\psi)}}{2r^2\sin\theta}
		\partial_{\theta}
		\left[e^{\lambda-\psi}r^{2}\sin^{3}\theta\frac{\partial\omega}{\partial \theta}\right]=\left(8\pi+2\beta\right)T^{t}_{\phi}.
	\end{split}
\end{equation}
Combining equations \ref{eq:17} and \ref{eq:19} results in
\begin{equation}\label{eq:20}
	\begin{split}
		-\frac{e^{-(\lambda+\psi)}}{2r^2\sin\theta}\partial_{r}\left[e^{-(\lambda+\psi)}r^{4}\sin^{3}\theta\frac{\partial\omega}{\partial r}\right]&-\frac{e^{-(\lambda+\psi)}}{2r^2\sin\theta}
		\partial_{\theta}
		\left[e^{\lambda-\psi}r^{2}\sin^{3}\theta\frac{\partial\omega}{\partial \theta}\right]\\&=\left(8\pi+2\beta\right)\left(\rho+p_{t}\right)e^{-2\psi}r^{2}\sin^{2}\theta\left(\Omega-\omega\right).
	\end{split}
\end{equation}	
Let define the angular velocity of fluid with respect to local frame $\bar{\omega}=\Omega-\omega$. Then, we can rewrite equation \ref{eq:20} as the following expression
\begin{equation}\label{eq:21}
	\begin{split}
		\frac{e^{\psi-\lambda}}{r^4}\partial_{r}\left[e^{-(\lambda+\psi)}r^{4}\frac{\partial\bar{\omega}}{\partial r}\right]+\frac{1}{r^2\sin^{3}\theta}
		\partial_{\theta}
		\left[\sin^{3}\theta\frac{\partial\bar{\omega}}{\partial \theta}\right]=4\left(4\pi+\beta\right)\left(\rho+p_{t}\right)\bar{\omega}.
	\end{split}
\end{equation}	
Angular velocity $\bar{\omega}(r,\theta)$ can be expanded in term of Legendre polynomial
\begin{equation}\label{eq:22}
	\bar{\omega}(r,\theta)=\sum_{l=1}^{\infty} \bar{\omega}_{l}(r)\left(-\frac{1}{\sin\theta}\frac{dP_{l}}{d\theta}\right),
\end{equation}
and $P_{l}$ satisfies Legendre differential equation
\begin{equation}\label{eq:23}
	\frac{d^{2}P_{l}}{d\theta^2}+\frac{\cos\theta}{\sin\theta}\frac{dP_{l}}{d\theta}+l(l+1)P_{l}=0.
\end{equation}

Furthermore, equation \ref{eq:21} can be reexpressed as
\begin{equation}\label{eq:24}
	\begin{split}
		\frac{e^{\psi-\lambda}}{r^4}\frac{d}{dr}\left[e^{-(\lambda+\psi)}r^{4}\frac{d\bar{\omega}_{l}}{dr}\right]-\frac{l(l+1)-2}{r^2}\bar{\omega}_{l}=
		4\left(4\pi+\beta\right)\left(\rho+p_{t}\right)\bar{\omega}_{l}.
	\end{split}
\end{equation}	
Considering the solution of $\bar{\omega}_{l}$ outside the star, we have to set $\rho=p_{t}=0$ and therefore the above equation transforms as
\begin{equation}\label{eq:25}
	\frac{d}{dr}\left[r^4\frac{d\bar{\omega}_{l}}{dr}\right]-\left[l(l+1)-2\right]\bar{\omega}_{l}r^2=0.
\end{equation}
To obtain explicit form of $\bar{\omega}$, we take the following ansatz
\begin{equation}\label{eq:26}
	\begin{split}
		\bar{\omega}_{l}&\propto r^{s}\\
		\bar{\omega}_{l}&\propto sr^{s-1}\\
		\bar{\omega}_{l}&\propto s(s-1)r^{s-2}.
	\end{split}
\end{equation}
The solution of $\bar{\omega}_{l}$ is $\bar{\omega}_{l}=d_{1}r^{l-1}+d_{2}r^{-(l+2)}$. However, due to the boundary condition, $l\geq2$ should be abandoned and the acceptable solution is only $l=1$. Consequently, we can write $\bar{\omega}_{l}$ is independent of $\theta$.

Interistingly, in the slow rotating approximation, the metric of neutron star is the same as that of Kerr metric. The $g_{t\phi}$ component of Kerr metric in the limit $(a<<r)$ can be written as
\begin{equation}\label{eq:27}
	g_{t\phi}^{\mathrm{Kerr}}\approx-\frac{2Ma\sin^{\theta}}{r}=-\frac{2J\sin^{2}\theta}{r},
\end{equation} 
where $J$ is the angular momentum. Equating this approximated expression with that of NS, it folows that $\omega=2J/r^3$ and then we find $\bar{\omega}=\Omega-2J/r^3$. As a result, equation \ref{eq:23} can be reformulated as
\begin{equation}\label{eq:28}
	\begin{split}
		\frac{e^{\psi-\lambda}}{r^4}\frac{d}{dr}\left[e^{-(\lambda+\psi)}r^{4}\frac{d\bar{\omega}_{l}}{dr}\right]&=
		4\left(4\pi+\beta\right)\left(\rho+p_{t}\right)\bar{\omega}_{l}\\
		\frac{d}{dr}\left[e^{-(\lambda+\psi)}r^{4}\frac{d\bar{\omega}_{l}}{dr}\right]&=
		4\left(4\pi+\beta\right)\left(\rho+p_{t}\right)\bar{\omega}_{l}r^{4}e^{\lambda-\psi}.
	\end{split}
\end{equation}
Integrating above equation from $r=0$ to the surface of the star $R$, we arrive at the following expression
\begin{equation}\label{eq:29}
	\begin{split}
		R^4\left.\frac{d\bar{\omega}}{dr}\right|_{r=R}&=4\left(4\pi+\beta\right)
		\int_{0}^{R} \left(\rho+p_r+\sigma\right)e^{\lambda-\psi}r^{4}\bar{\omega} \, dr\\
		J&=\frac{2}{3}\left(4\pi+\beta\right)
		\int_{0}^{R} \left(\rho+p_r+\sigma\right)e^{\lambda-\psi}r^{4}\bar{\omega} \, dr
	\end{split}
\end{equation}
It has been know that moment of inertia can $I$ and $J$ are linked by the relation $I=J/\Omega$. Thus, moment of inertia can be calculated as
\begin{equation}\label{eq:30}
	I=\frac{2}{3}\left(4\pi+\beta\right)
	\int_{0}^{R} \left(\rho+p_r+\sigma\right)e^{\lambda-\psi}r^{4}\frac{\bar{\omega}}{\Omega} \, dr.
\end{equation}

\section{Tidal love numbers in $f(R,T)$ gravity}\label{sec5}
The tidal love numbers of neutron stars are investigated by doing metric perturbations of the static solutions in $f(R,T)$ gravity. According to Regge-Wheeler formalism, such perturbations can be divided into two types namely even and odd perturbations resulting in polar and axial love numbers, respectively. In this section, we will derive the analytical solutions of love numbers for both cases. In order to do so, we need to obtain perturbed field equations.  
Taking variation of equation \ref{eq:18} yields
\begin{equation}\label{eq:31}
		\delta G^{\alpha}_{\nu}=\left(8\pi+2\beta\right)\delta T^{\alpha}_{\nu}+\beta\left(-\delta\rho+\frac{1}{3}\delta p_r+\frac{2}{3}\delta p_t\right)\delta^{\alpha}_{\nu}.
\end{equation}

We focus on a static NS that is perturbed by an external tidal field. Moreover, the tidal field is assumed to be weak and time-independent. Due to this field, the multipole structure inside the star is generated. Geometrically, the tidal field deforms the spacetime metric outside the star. The deformed metric is given by
\begin{equation}\label{eq:32}
	g_{\mu\nu}=g_{\mu\nu}^{(0)}+h_{\mu\nu},
\end{equation} 
where $g_{\mu\nu}^{(0)}$ is the background or unperturbed metric and $h_{\mu\nu}$ is a small perturbation. Based on parity under rotation, the metric $h_{\mu\nu}$ can be decomposed as
\begin{equation}\label{eq:33}
	h_{\mu\nu}=h_{\mu\nu}^{\mathrm{even}}+h_{\mu\nu}^{\mathrm{odd}}.
\end{equation}

\subsection{Polar love number}
The polar love number is calculated using polar or even perturbation. This type of love number is also called electric love number. In Regge-Wheeler gauge, the metric $h_{\mu\nu}^{\mathrm{even}}$ can be expressed as
\begin{equation}\label{eq:34}
	h_{\mu\nu}^{\mathrm{even}}=
	\left[\begin{array}{cccc}
		-e^{2\psi}H_{0} & H_{1} & 0 & 0	\\
		H_{1} & e^{2\lambda}H_2 & 0 & 0	\\
		0 & 0 & Kr^2 & 0	\\
		0 & 0 & 0 & Kr^2\sin^{2}\theta
	\end{array}\right]Y^{m}_{l}\left(\theta,\phi\right).
\end{equation}
The components of perturbed Einstein tensor are given in Appendix.
In this work, we consider quadrupolar love number and therefore we take $l=2$. 
From equations \ref{eq:57}, \ref{eq:58}, and \ref{eq:59}, we obtain the following relations
\begin{equation}\label{eq:35}
	\begin{split}
		H_{0}&=H_{2}=H\\
		K'&=H'+2\psi'H\\
		k''&=H''+2\psi''H+2\psi'H'.
	\end{split}
\end{equation}

Subtracting $tt$ from $rr$-components of perturbed Einstein tensor and using equation \ref{eq:18} yields
\begin{equation}\label{eq:36}
	\delta G^{t}_{t}-\delta G^{r}_{r}=-\left(8\pi+2\beta\right)\left(1+\frac{\delta p_r}{\delta\rho}\right)\delta \rho.
\end{equation}
Putting equations \ref{eq:60} and \ref{eq:61} into \ref{eq:36} gives
\begin{equation}\label{eq:37}
	\begin{split}
		\left[-e^{-2\lambda}K''+\frac{K'}{r}e^{-2\lambda}\left[-2+(\lambda'+\psi')r\right]+\frac{H}{r}\left[-2r\lambda'-2r\psi'\right]\right]Y^{0}_{2}\\=-\left(8\pi+2\beta\right)\left(1+\frac{\delta p_r}{\delta\rho}\right)\delta \rho.
	\end{split}
\end{equation}
Combining equations \ref{eq:31}, \ref{eq:35}, and \ref{eq:62} yields
\begin{equation}\label{eq:38}
	\begin{split}
		\frac{\delta G^{\theta}_{\theta}+\delta G^{\phi}_{\phi}}{2Y^{0}_{2}}&=-e^{-2\lambda}\frac{\left(\psi'+\lambda'\right)}{r}H\\
		\frac{\left[8\pi\frac{\delta p_{t}}{\delta\rho}+\beta\left(-1+\frac{1}{3}\frac{\delta p_r}{\delta\rho}+\frac{8}{3}\frac{\delta p_t}{\delta\rho}\right)\right]}{Y^{0}_{2}}\delta\rho&=-e^{-2\lambda}\frac{\left(\psi'+\lambda'\right)}{r}H.
	\end{split}
\end{equation}
Combining equations \ref{eq:35} and \ref{eq:37}, we get
\begin{equation}\label{eq:39}
	\begin{split}
		H''+\left[\frac{2}{r}+\psi'-\lambda'\right]H'&+\left[2e^{-2\lambda}\psi''-\frac{6}{r^2}+e^{-2\lambda}\left(\frac{6\psi'+2\lambda'}{r}-2\lambda'\psi'-2\psi'^2\right)\right]H\\
		&+\left[\frac{\left(8\pi+2\beta\right)\left(1+\frac{\delta p_r}{\delta\rho}\right)}{8\pi\frac{\delta p_t}{\delta\rho}+\beta\left(-1+\frac{1}{3}\frac{\delta p_r}{\delta\rho}+\frac{8}{3}\frac{\delta p_t}{\delta\rho}\right)}
		\frac{\psi'+\lambda'}{r}\right]H=0.
	\end{split}
\end{equation}
Outside the star, $\rho$, $p_{t}$, $p_{r}$ and their derivatives are zero. Consequently, differential equation for $H$ in this region can be simplified as
\begin{equation}\label{eq:40}
	H''+\frac{2\left(r-M\right)}{r\left(r-2M\right)}H'-\frac{4M^2-12Mr+6r^2}{r^2\left(r-2M\right)^2}H=0.
\end{equation}

Making change of variable $x=r/M-1$, equation \ref{eq:40} transforms to Legendre differential equation
\begin{equation}\label{eq:41}
\left(1-x^2\right)H''-2xH'+\left(6-\frac{4}{1-x^2}\right)H=0.
\end{equation}
The solution of $H$ outside the star is 
\begin{equation}\label{eq:42}
	H(r)=c_{1}P^{2}_{2}\left(\frac{r}{M}-1\right)+c_{2}Q^{2}_{2}\left(\frac{r}{M}-1\right),
\end{equation}
and the asymtotic form of $H(r)$ can be approximated as
\begin{equation}\label{eq:43}
	H(r)=\frac{8}{5}c_{1}\frac{M^3}{r^3}+O\left(\frac{M^4}{r^4}\right)+3c_{2}\frac{r^2}{M^2}+O\left(\frac{r}{M}\right).
\end{equation}
Polar tidal love number $k_{2}^{\mathrm{polar}}$ and tidal deformability $\Lambda$ are defined as
\begin{equation}\label{eq:44}
	\begin{split}
		k^{E}_{\mathrm{polar}}&=\frac{1}{2R^5}\frac{c_{1_{\infty}}}{c_{2_{\infty}}}=\frac{4}{15}C^5\frac{c_{1}}{c_{2}}\\
		k^{E}_{\mathrm{polar}}&=\frac{8}{5}C^5\left(1-2C\right)^{2}\left[2+2C\left(y-1\right)-y\right]\\ 
		&\times
		\left\{
		\begin{aligned}
			2C\left[6-3y+3C\left(5y-8\right)\right]+4C^{3}\left[13-11y+C\left(3y-2\right)+2C^{2}\left(1+y\right)\right]\\
			+3\left(1-2C\right)^{2}\left[2-y+2C\left(y-1\right)\right]\ln\left(1-2C\right)
		\end{aligned}
		\right\}^{-1}\\
		\Lambda&=\frac{2}{3}\frac{k^{E}_{\mathrm{polar}}}{C^5}.
	\end{split}
\end{equation}
Coefficients $c_{1_{\infty}}$ and $c_{2_{\infty}}$ are the coefficients of decaying and growing parts of the asymtotic form of $H$, respectively. In addition, $C=M/R$ is the compactness of the star and $y$ is given by $y=RH'(R)/H(R)$.

\subsection{Axial love number}
The axial love number is calculated using axial or odd perturbation. This type of love number is also called magnetic love number. In Regge-Wheeler gauge, the metric $h_{\mu\nu}^{\mathrm{odd}}$ can be expressed as
\begin{equation}\label{eq:45}
	h_{\mu\nu}^{\mathrm{even}}=
	\left[\begin{array}{cccc}
		0 & 0 & h_{0}S^{ml}_{\theta} & h_{0}S^{ml}_{\phi}	\\
		0 & 0 & h_{1}S^{ml}_{\theta} & h_{1}S^{ml}_{\phi}	\\
		h_{0}S^{ml}_{\theta} & h_{1}S^{ml}_{\theta} & 0 & 0	\\
		h_{0}S^{ml}_{\phi} & h_{1}S^{ml}_{\phi} & 0 & 0
	\end{array}\right],
\end{equation}
where $S^{ml}_{\theta}=-\frac{1}{\sin\theta} \frac{dY^{m}_{l}}{d\phi}$
and $S^{ml}_{\phi}=\sin\theta \frac{dY^{m}_{l}}{d\theta}$.
The $\theta\phi$ and $t\phi$-components of perturbed mixed tensor field are
\begin{equation}\label{eq:46}
	\begin{split}
		\delta R^{\theta}_{\phi}-\frac{1}{2}\delta^{\theta}_{\phi}R&=\left(8\pi+2\beta\right)T^{\theta}_{\phi}+\delta^{\theta}_{\phi}\left[\delta T-2\delta\mathcal{L}_{m}\right]\\
		-e^{-\lambda-\psi}\partial_{r}\left(e^{\psi-\lambda}h_{1}\right)&=0\rightarrow h_{1}=0.
	\end{split}
\end{equation}
and
\begin{equation}\label{eq:47}
	\begin{split}
		\delta R^{t}_{\phi}-\frac{1}{2}\delta^{t}_{\phi}R&=\left(8\pi+2\beta\right)T^{t}_{\phi}+\delta^{t}_{\phi}\left[\delta T-2\delta\mathcal{L}_{m}\right]\\
		h''_{0}-\left(\lambda'+\psi'\right)h'_{0}&+\left\{-\frac{(l-1)(l+2)}{r^2}e^{2\lambda}+\frac{2(\psi'+\lambda')}{r}-\frac{2}{r^2}\right\}h_{0}=0.
	\end{split}
\end{equation}
We assume that $h_{0}$ is time-independent and putting $l=2$ into above equation, we obtain
\begin{equation}\label{eq:48}
	h''_{0}-\left(\lambda'+\psi'\right)h'_{0}+\left\{-\frac{4}{r^2}e^{2\lambda}+\frac{2(\psi'+\lambda')}{r}-\frac{2}{r^2}\right\}h_{0}=0.
\end{equation}
Outside the star, the same conditions in the previous subsection applies here as well. Hence, the differential equation of $h_{0}$ becomes
\begin{equation}\label{eq:49}
	h''_{0}+\left[\frac{4M-6r}{r^2\left(r-2M\right)}\right]h_{0}=0
\end{equation}
Let us consider change of variable $x=2M/r$. Consequently, equation \ref{eq:49} can be rewritten as
\begin{equation}\label{eq:50}
	x(x-1)h''_{0}+\frac{\left(1-x\right)}{2}h'_{0}+\left(-\frac{6}{x}+2\right)h_{0}=0.
\end{equation}
Moreover, the solution of equation \ref{eq:50} is
\begin{equation}\label{eq:51}
	h_{0}\left(r\right)=b_{1}\left(\frac{r}{2M}\right)^{2}{}_2F_1(-1,4;4;\frac{r}{2M})+b_{1}\left(\frac{2M}{r}\right)^{2}{}_2F_1(1,4;6;\frac{2M}{r}),
\end{equation}
while its asymtotic form corresponds to
\begin{equation}\label{eq:52}
	h_{0}\left(r\right)=-\frac{1}{5}b_{1}\frac{M^2}{r^2}+O\left(\frac{M^3}{r^3}\right)+b_{2}\left(\frac{r}{M}\right)^{3}+O\left(\frac{r^2}{M^2}\right).
\end{equation}
The axial love number is defined as
\begin{equation}\label{eq:53}
	k^{\mathrm{axial}}_{2}=-\frac{1}{3}\frac{1}{R^5}\frac{b_{1_{\infty}}}{b_{2_{\infty}}}
	=\frac{1}{15}C^{5}\frac{b_{1}}{b_{2}},
\end{equation}
where $b_{1_{\infty}}$ and $b_{2_{\infty}}$ stand for the coefficients of decaying and growing form of the asymtotic form of $h_{0}$. 
We arrive at the axial tidal love number of neutron star
\begin{equation}\label{eq:54}
	\begin{split}
		k^{\mathrm{axial}}_{2}=&\frac{8}{5}C^{5}\left[2C\left(y-2\right)-y+3\right]\\
		&\times
		\left\{
		\begin{aligned}
		2C\left[2C^{3}(y+1)+2C^{2}y+3C(y-1)-3y+9\right]\\+3C\left[2C(y-2)-y+3\right]\ln(1-2C)
	\end{aligned}
	\right\}^{-1}		
	\end{split}.
\end{equation}
Here, $y$ is defined as $y=Rh_{0}'(R)/h_{0}(R)$.
\section{Numerical results}

\subsection{Mass-radius relation}
\begin{figure}[H]
	\centering
	\includegraphics[width=1.0\textwidth]{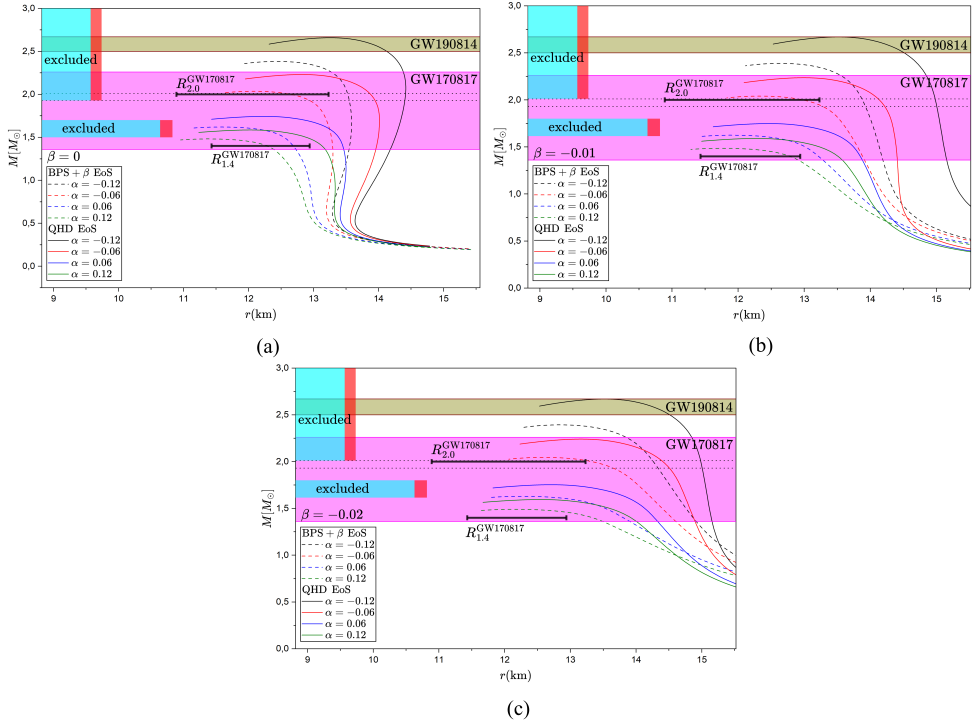}
	\caption{Mass of Neutron stars vs radius for two different EoS plotted based on variations of $\beta$ and $\alpha$, where (a)$\beta=0\hspace{1mm}\mathrm{(GR)}$, (b) $\beta=-0.01$, and (c) $\beta=-0.02$. The vertical and horizontal cyan+red regions in the $M$ vs $r$ curve are the additional constraints from $\mathrm{GW170817}$ showing areas that cannot be occupied by neutron stars (excluded regions). The horizontal cyan+red regions indicate maximum masses that are too small for neutron stars. Meanwhile, the vertical cyan+red regions indicate that stars within these regions have radii that are too small.\cite{bauswein2017neutron}The two horizontal black lines are the allowed radii for neutron  stars according to $\mathrm{GW170817}$. The allowed radii of NSs with masses of $1.4\hspace{1mm}M_{\odot}$  and $2.0\hspace{1mm}M_{\odot}$ are  $R_{1.4}=12.42_{-0.99}^{+0.52}\hspace{1mm}\mathrm{km}$ and $R_{2.0}=12.12_{-1.23}^{+1.11}\hspace{1mm}\mathrm{km}$.\cite{altiparmak2022sound}.}
	\label{fig1}
\end{figure}
Figure \ref{fig1} represents the plots of mass as a function of radius varied with respect to the free parameters $\beta$ and anisotropy $\alpha$. It can be seen that anistropy parameter $\alpha$ significantly influences the maximum achievable mass of NSs. The more negative $\alpha$ gives the higher mass, because the negative anisotropy $\left(p_t-p_r<0\right)$ introduces an additional force into the hydrostatic equilibrium equations. The condition $p_r>p_t$ resists gravitational collapse more strongly than in isotropic stars, allowing the star to accommodate more mass. Furthermore, the negative $\alpha$ also increases the compactness of NSs shown by the curves that for the same radius, NSs with more negative $\alpha$ have higher masses.

A similar thing applies to the parameter $\beta$, where the more negative $\beta$ produces the higher mass for the same anisotropy parameter $\alpha$. However, its impact on mass increment is not as large as that of $\alpha$. In constrast to $\alpha$, the negative $\beta$ reduces the compactness of NSs quite significantly because NSs in $f(R,T)$ gravity with the same mass as GR have larger radii for negative $\beta$.  We can conclude that the coupling between geometry and matter has an opposite effect on the mass and compactness of NSs.

We use three values of $\beta$ namely $\beta=0$, $\beta=-0.01$, and $\beta=-0.02$. For anisotropic parameter $\alpha$, we examine four cases of $\alpha$: $\alpha=-0.12$, $\alpha=-0.06$, $\alpha=0.06$, and $\alpha=0.12$. As demonstrated in figure \ref{fig1}, the mass constraint of GW170817 can be achieved by NSs described by both QHD and $\mathrm{BPS}+\beta$ EoS with all values of $\beta$ and $\alpha$. Furthermore, all curves successfully avoid excluded areas from GW170817 constraint. However, for $\mathrm{BPS}+\beta$ EoS with $\beta=0$, we have $\alpha=0.12$ and $\alpha=0.06$ that satisfy $R_{1.4}^{\mathrm{GW170817}}$ constraint while $\alpha=-0.06$ fullfill $R_{2.0}^{\mathrm{GW170817}}$ constraint. In the case of $\beta=-0.01$, there are two curves i.e. $\alpha=0.12$ and $\alpha=-0.06$ that obey $R_{1.4}^{\mathrm{GW170817}}$ $R_{2.0}^{\mathrm{GW170817}}$ constraints, respectively. Interestingly, there are only three curves that can achieve the mass constraint of GW190814. This can only be attained by NSs filled with QHD EoS for $\alpha=-0.12$ and all values of $\beta$. Interestingly, only for $\beta=-0.02$ NSs can attain the upper bound of the mass constraint of GW 190814 i.e. $2.67\hspace{1mm}M_{\odot}$. From the plots, we can also infer that for the same radius, NSs governed with QHD EoS have higher masses compare to $\mathrm{BPS}+\beta$ meaning that NSs filled with QHD Eos are more compact.

\subsection{Moment of inertia}
\begin{figure}[H]
	\centering
	\includegraphics[width=1.0\textwidth]{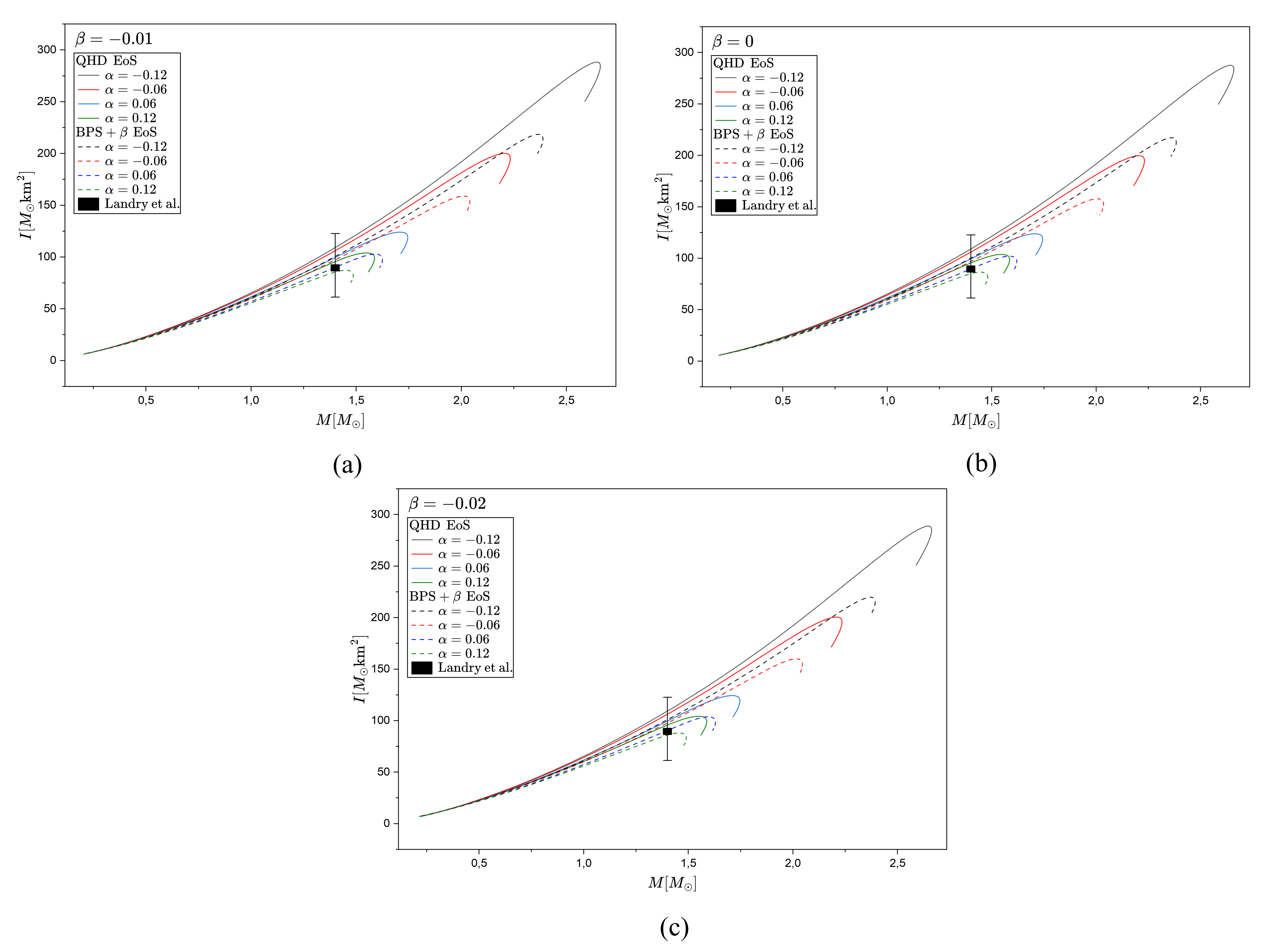}
	\caption{Moment of inertia vs radius for two different EoS plotted based on variations of $\beta$ and $\alpha$, where (a)$\beta=0\hspace{1mm}\mathrm{(GR)}$, (b) $\beta=-0.01$, and (c) $\beta=-0.02$. }
	\label{fig2}
\end{figure}
Another quantity that is strongly related to the mass of NSs is the moment of inertia. The plots of it for three different $\beta$ are quite similar and the impact of parameters $\alpha$ and $\beta$ on moment of inertia is the same as in the mass plot where negative $\alpha$ magnifies $I$ significantly while negative $\beta$ has slight increment effect of $I$. To justify the correct plots, we employ the constraint of moment of inertia provided by Ref. \citep{landry2020nonparametric}. This constraint is determined from radio observation of heavy pulsar where the lower limits on the maximum mass of NSs are taken. From figure \ref{fig2}, it is obvious that all results satisfy the constraint. Nevertheless, the most favorable outcome is evaluated for $\beta=-0.02$ where we have two curves that intersects with the central value of constraint. 

\subsection{Polar tidal love number}\label{sec6.3}
\begin{figure}[H]
	\centering
	\includegraphics[width=1.0\textwidth]{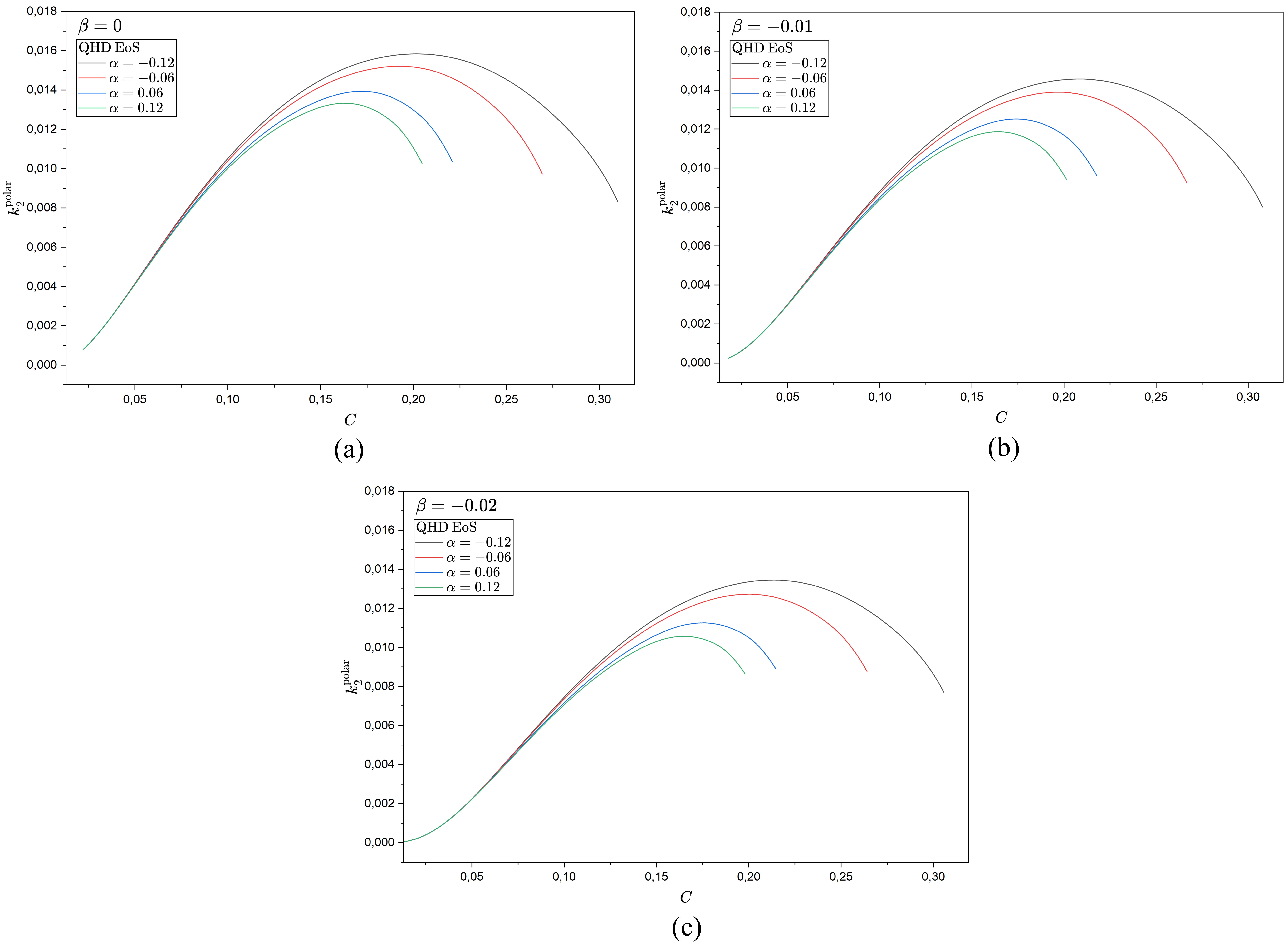}
	\caption{Polar TLNs vs compactness for QHD EoS plotted based on variations of $\beta$ and $\alpha$, where (a)$\beta=0\hspace{1mm}\mathrm{(GR)}$, (b) $\beta=-0.01$, and (c) $\beta=-0.02$.}
	\label{fig3}
\end{figure}
With regards to the polar TLNs, we divide the plots of two EoS into two figures due to the difference between both peaks are too large. As illustrated in figure \ref{fig3}, QHD EoS is extremely stiff shown by the low values of the polar TLNs. Initially, the curves rise drastically as the compactness increases until reach their maximum values. After this, they decreases dramatically. This indicates that a NS with small compactness is difficult to deformed because the internal gravitational binding is too weak to unify the response of the entire NS. In addition, there is a value of compactness where the polar TLNs reach their peaks implying that at this point the internal binding of a NS is optimally balanced but it is not strong enough to prevent the tidal deformation. After that point, the compactness of the NS rises and the NS becomes more stable. As a result, the polar TLNs tends to zero.

\begin{figure}[H]
	\centering
	\includegraphics[width=1.0\textwidth]{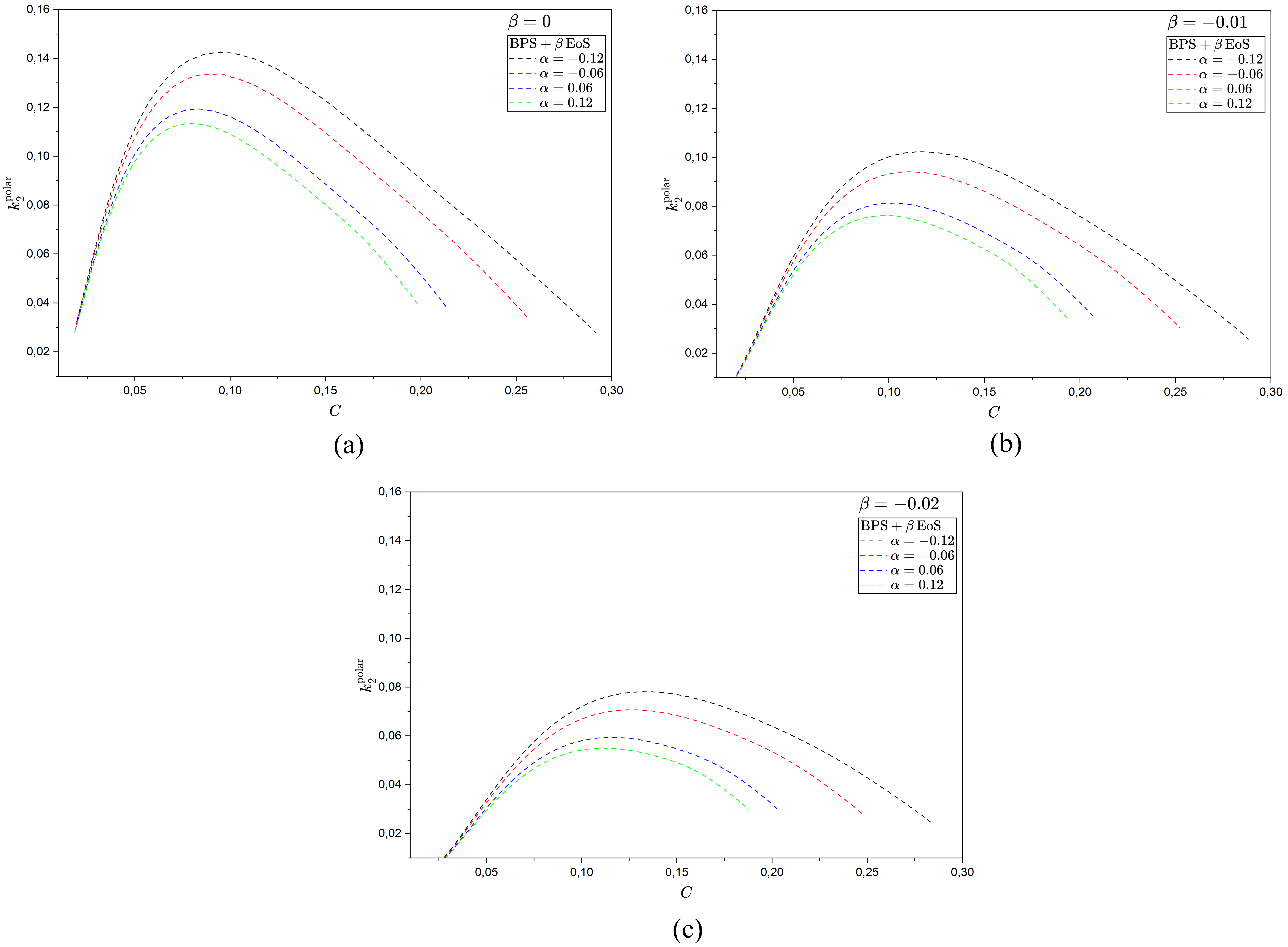}
	\caption{Polar TLNs vs compactness for BPS EoS plotted based on variations of $\beta$ and $\alpha$, where (a)$\beta=0\hspace{1mm}\mathrm{(GR)}$, (b) $\beta=-0.01$, and (c) $\beta=-0.02$.}
	\label{fig4}
\end{figure}
We also deduce that the negative $\alpha$ increases the polar deformation of NSs with the same compactness. This suggests that if there are NSs with the same compactness, then the NSs that are more easily deformed are the ones that have more negative anisotropy or a greater difference between radial and tangential pressure.
Conversely, parameter $\beta$ has the opposite effect to $\alpha$ on the polar TLNs where the negative $\beta$ decreases the polar TLNs quite more significantly. Thus, the negative coupling between matter and geometry can help NSs to minimize the polar deformation. In addition, the values of compactness which results in maximum TLNs are shifted to the right when negative $\beta$ is employed. 

The features of TLNs vs compactness curves depicted in figure \ref{fig4} are the same as in figure \ref{fig3}. Nevertheless, the TLNs of NSs filled with $\mathrm{BPS}+\beta$ EoS are larger than QHD. This reveals that QHD EoS is much more stiff than $\mathrm{BPS}+\beta$ EoS and consequently the NSs with this EoS is much more robust. Additionally, the impact of $\beta$ on minimizing the TLNs is more remarkable which means that in this case the deformation of NSs is quite sensitive to the coupling between geometry and matter. Although the maximum TLNs in figure \ref{fig4} are higher than figure \ref{fig3}, the points of compactness giving the maximum TLNs are less. We find that there are drastic increase in the curves at small compactness and after achieve the highest point these curves decrase linearly to zero. 

It is worthwhile to compare our results with Ref. \citep{yazadjiev2018tidal} to emphasize the role of MG although the anisotropy is not considered in that paper. In that reference, the polar TLNs in $f(R)$ gravity formulation is higher than in GR due to the existence of scalar field that can prevent deformation. In contrast, our results depict that for negative $\beta$ the polar TLNs in $f(R,T)$ gravity can be reduced to be lower than that of GR. Therefore, the $2\beta T$ term in this research has the opposite role to the $R^2$ term in Ref. \citep{yazadjiev2018tidal}. For this reason, it is clear that MG has a significant impact on the internal geometry of NSs spacetime.

\subsection{Axial tidal love number}\label{sec6.4}
\begin{figure}[H]
	\centering
	\includegraphics[width=1.0\textwidth]{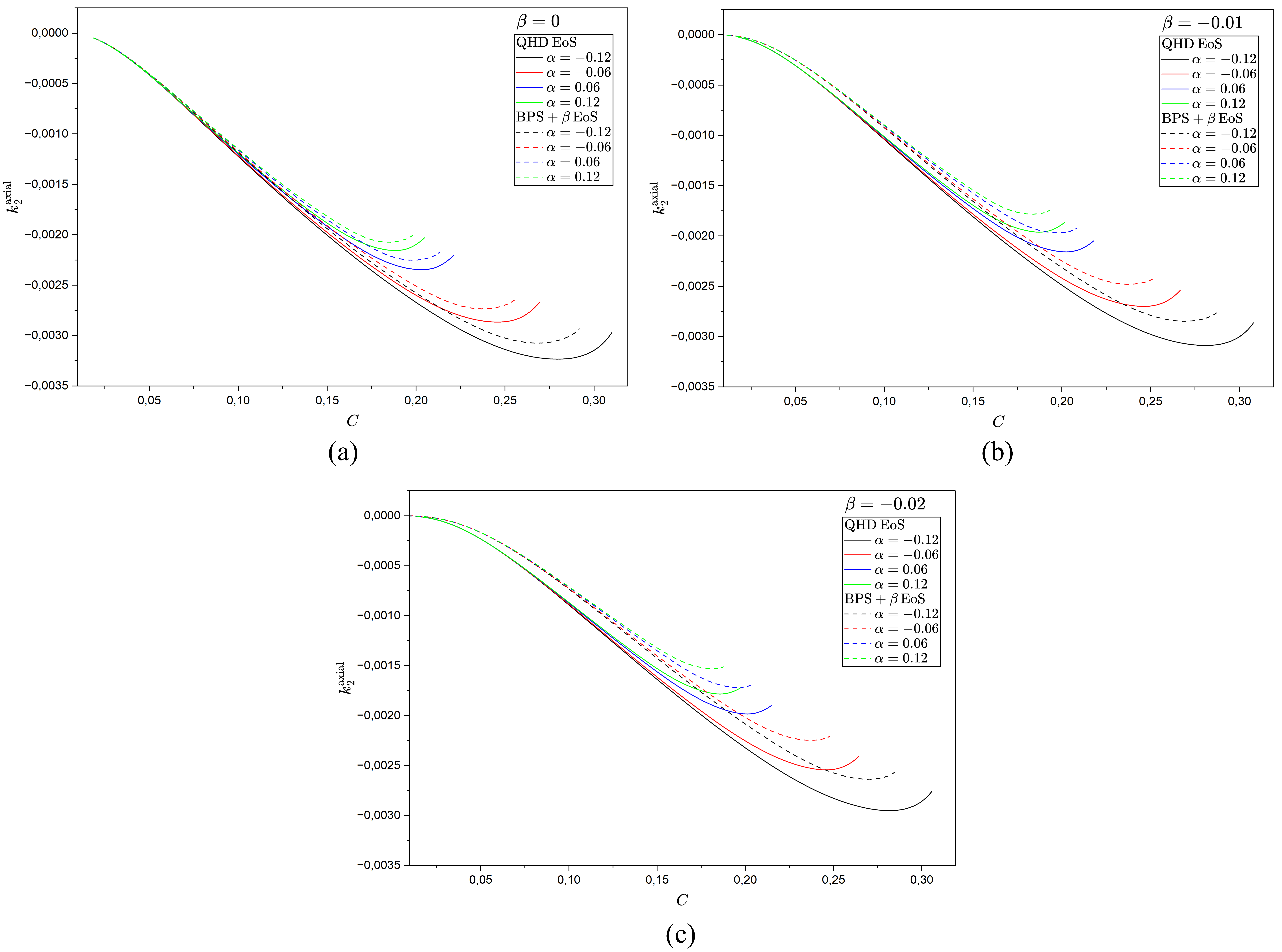}
	\caption{Axial TLNs vs compactness for two different EoS plotted based on variations of $\beta$ and $\alpha$, where (a)$\beta=0\hspace{1mm}\mathrm{(GR)}$, (b) $\beta=-0.01$, and (c) $\beta=-0.02$.}
	\label{fig5}
\end{figure}
The axial TLNs, in contrast, have different features compared to the polar TLNs. As presented in figure \ref{fig5}, the negative $\beta$ decreases the absolute values axial TLNs while the negative $\alpha$ increases them. Interestingly, from figure \ref{fig5} we can see clearly that NSs with QHD EoS is easier to be axially deformed than those with $\mathrm{BPS}+\beta$ EoS. By constrast to the polar TLNs, the axial TLNs for both EoS do not differ too much meaning that they have the same order of magnitude. Unfortunately, the axial TLNs are extremely small which make them difficult to be detected by detectors. Practically, in GR the axial TLNs are usually neglected in the observation of GWs because they do not influence the phase change of GWs signal\citep{jimenez2018impact}. It implies that our results matches the common asumption used in the GWs observation.

\subsection{Tidal deformability}\label{sec6.5}
\begin{figure}[H]
	\centering
	\includegraphics[width=1.0\textwidth]{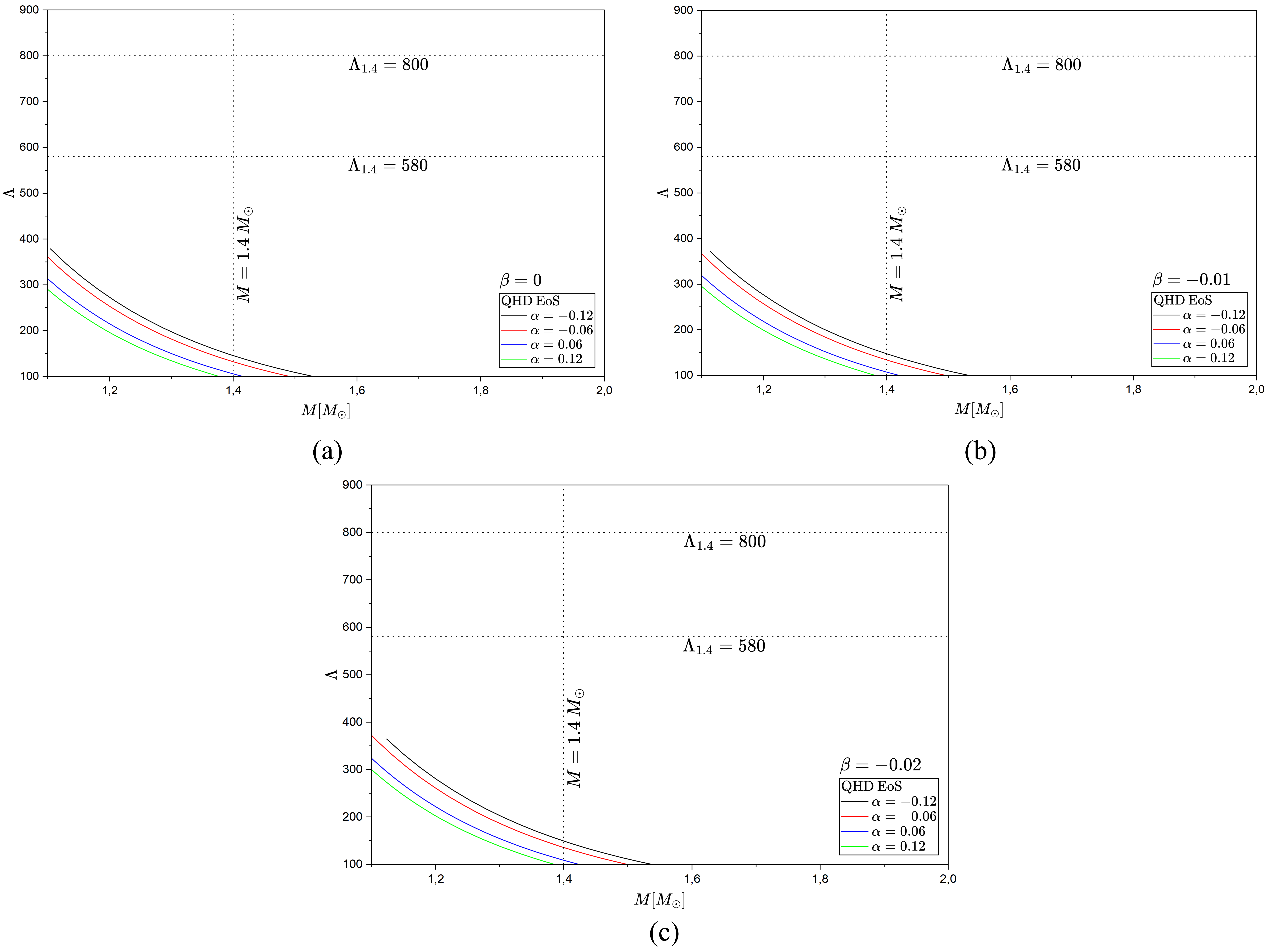}
	\caption{Tidal deformability vs mass for QHD EoS plotted based on variations of $\beta$ and $\alpha$, where (a)$\beta=0\hspace{1mm}\mathrm{(GR)}$, (b) $\beta=-0.01$, and (c) $\beta=-0.02$. }
	\label{fig6}
\end{figure}
The tidal deformability $\left(\Lambda\right)$ is calculated based on the polar TLNs as given by equation \ref{eq:44}. As we have done in the subsection \ref{sec6.3}, in this subsection separate the plots of $\Lambda$ vs $M$ for both equations of state to provide better visibility. The curves of $\Lambda$ for three different $\beta$ shown in figure \ref{fig6} are nearly similar. The negative $\alpha$ and $\beta$ gives higher values of $\Lambda$ but actually the effect of $\beta$ is not too significant compared to $\alpha$. As the mass increases $\Lambda$ decreases gradually. When $M=1.4\hspace{1mm} M_{\odot}$, $\Lambda$ is constrained to be less than $200$.

Due to the proportionality between $k_{2}^{\mathrm{polar}}$ and $\Lambda$, the values of $\Lambda$ are axtremely small and difficult to be detected in GWs observation. Then, these values are
much smaller than GW170817 constraint. Consequently, a NS with this property is not appropriate to be NSs candidate for GW 170817. However, recent observation of GWs given by GW 190814 found a compact object of mass between $2.50\hspace{1mm}M_{\odot}$ and $2.67\hspace{1mm}M_{\odot}$. In fact, there is no $\Lambda$ constraint of that object and it is still a matter of debate among physicists today whether the object is a black hole or a NS. If the object is a NS, then the possible reason for the lack of $\Lambda$ data is that its value is too small to be observed.
It can therefore be inferred that, a NS with QHD EoS can still become candidate for secondary object of GW190814.

\begin{figure}[H]
	\centering
	\includegraphics[width=1.0\textwidth]{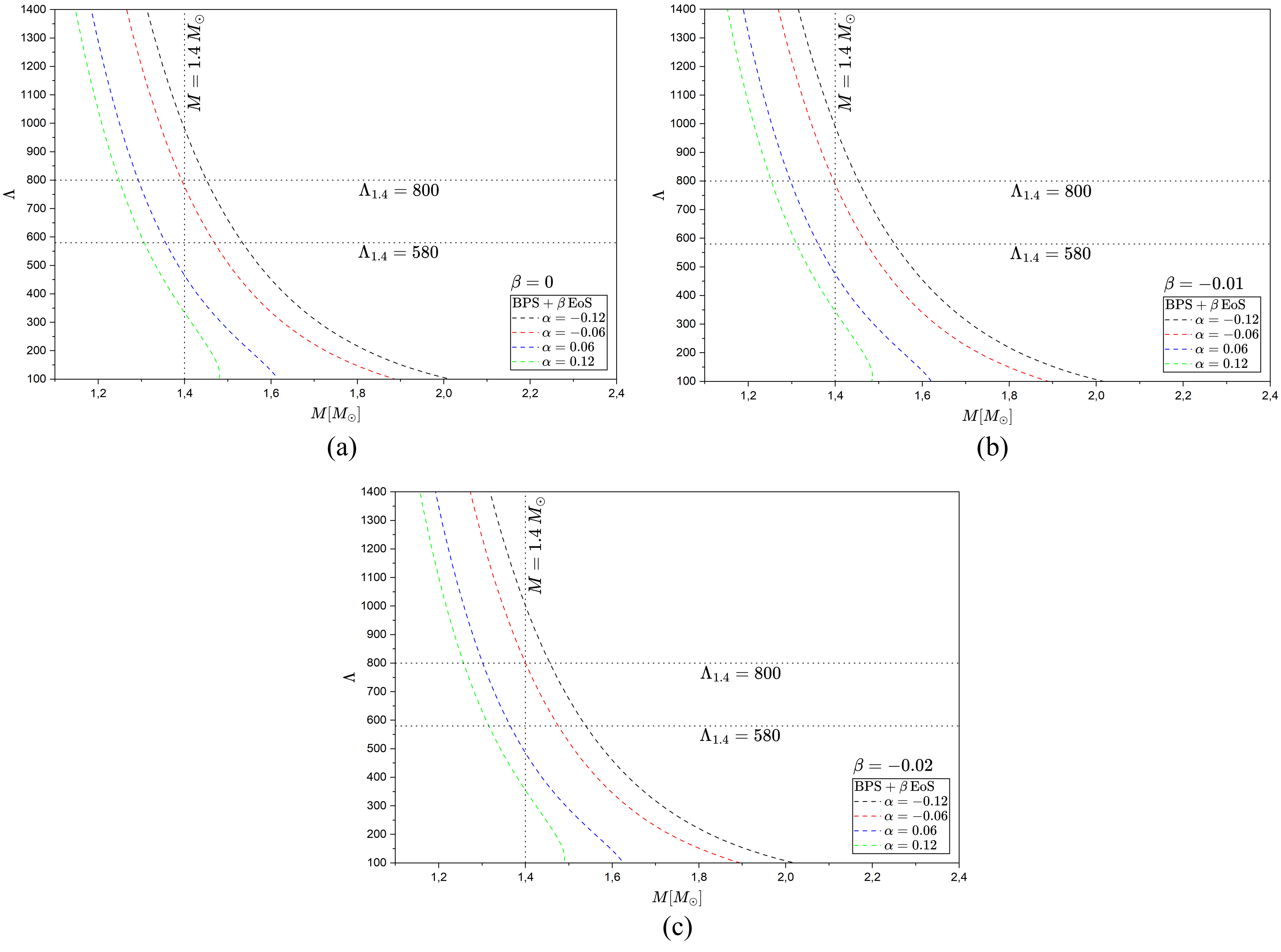}
	\caption{Tidal deformability vs mass for $\mathrm{BPS}+\beta$ EoS plotted based on variations of $\beta$ and $\alpha$, where (a)$\beta=0\hspace{1mm}\mathrm{(GR)}$, (b) $\beta=-0.01$, and (c) $\beta=-0.02$.}
	\label{fig7}
\end{figure}
The features of $\Lambda$ for NSs with  $\mathrm{BPS}+\beta$ EoS are similar to that of NSs in the previous analysis. Nevertheless, the values of $\Lambda$ shown in figure \ref{fig7} are much higher. As we can see, these values are quite large enough to be detected by observatory. Interestingly, for three diferent $\beta$ there is only one value of $\alpha$ namely $\alpha=-0.06$ that satisfy GW170817 constraint. For $\beta=-0.02$ and $\alpha=-0.06$, the curve exactly intersects the point with coordinates $\left(1.4;580\right)$ which corresponds to the upper bound of that constraint. Based on these properties, a NS with $\mathrm{BPS}+\beta$ EoS can be a strong candidate for a NS according to GW170817 constraint.

\section{Discussion}
In this work, we have investigated the mass-radius relation, moment of inertia, and tidal love number of anisotropic NSs in the framework of $f(R,T)$ gravity especially $f(R,T)=R+2\beta T$ gravity. 
We use three values of $\beta$, i.e. $\beta=0$, $\beta=-0.01$, and $\beta=-0.02$.
We use anisotropic model given by Horvat \cite{horvat2010radial} and employ four diferent values of parameter $\alpha$: $\alpha=-0.12$, $\alpha=-0.06$, $\alpha=0.06$, and $\alpha=0.12$. Then, we consider two different equations of state to model NSs.  To constraint plots of the mass-radius relation and the tidal deformability, we adopt the constraints from GW170817 and GW190814. 

We find that free parameters $\alpha$ and $\beta$ influence the static mass of NSs where $\alpha$ has a more significant impact than $\beta$. By choosing some specific values of $\alpha$ and $\beta$, the NSs can achieve all given constraints. The NSs governed by $\mathrm{BPS+\beta}$ EoS can reach the GW170817 constraint for all $\alpha$ and $\beta$. 
Furthermore, the excluded areas from GW170817 constraint are successfully avoided. But, for $\beta=0$ we have $\alpha=0.12$ and $\alpha=0.06$ that obey $R_{1.4}^{\mathrm{GW170817}}$ constraint while $\alpha=-0.06$ fullfill $R_{2.0}^{\mathrm{GW170817}}$ constraint. In the case of $\beta=-0.01$, we obtain more specific results where there are two curves i.e. $\alpha=0.12$ and $\alpha=-0.06$ that obey $R_{1.4}^{\mathrm{GW170817}}$ $R_{2.0}^{\mathrm{GW170817}}$ constraints, respectively.
But, the NSs with $\alpha=-0.12$ and $\beta=-0.02$ has the highest maximum mass that nearly achieve the maximum constraint of GW170817. On the other hand, the NSs modeled using QHD EoS remain within GW170817 bound for all $\alpha$ and $\beta$. However, the ones that meet GW190814 constraint only those with $\alpha=-0.12$ and all $\beta$. Surprisingly, there is only one curve that can achieve the upper limit of GW190814 which corresponds to $\alpha=-0.12$ and $\beta=-0.02$.

When considering the moment of inertia, the NSs with all taken parameters are consistent with the observation constraint from Landry et al. \cite{landry2020nonparametric}. In this case, $\alpha$ has a more considerable effect to high value of moment inertia than $\beta$. Nevertheless, the NSs described by $\mathrm{BPS}+\beta$ EoS that intersect the central constraint. Meanwhile, the plots of the NSs composed of QHD EoS lie above it. As demonstrated the calculation static mass of NSs, the NSs governed by $\mathrm{BPS}+\beta$ EoS have larger the moment of inertia than the others.

The polar TLNs of the NSs constructed with the $\mathrm{BPS}+\beta$ EoS are considerably high while the others described by the QHD EoS are extremely low. Conversely, the axial TLNs for both types of NSs nearly have the same absolute values where the NSs with $\mathrm{BPS}+\beta$ EoS are slightly higher.  
The features of the tidal deformability $(\Lambda)$ are similar to the polar TLNs. The NSs with $\mathrm{BPS}+\beta$ EoS agree with the constraints imposed by the GW170817 for specific values of $\beta$ i.e., $\beta=0$, $\beta=-0.01$, $\beta=-0.02$ and $\alpha=-0.06$. Hence, the NSs having these features are viable candidates for a NS based on the GW170817 constraint. On the other hand, the NSs modeled with QHD EoS have remarkably low and are not compatible with GW170817 constraint. Nevertheless, the NSs with these properties and have masses in the range between $\left(2.50-2.67\right)M_{\odot}$ are still considered acceptable within the GW190814 bounds. In conclusion, we argue that the secondary object observed in GWs event GW190814 is a NS.

\backmatter

\bmhead{Acknowledgements}
We sincerely thank Prof. Luiz L. Lopes for sharing the EOS data.  Yusmantoro and M.L.P. would like to thank Indonesia Endowment Fund for Education Agency (Lembaga Pengelola Dana Pendidikan/LPDP). FPZ also wants to express many thanks all members of Theoretical Physics Laboratory, Institut Teknologi Bandung for the hospitality and useful discussion.

\section*{Declaration}
The authors declare no competing interests.

\section*{Appendix: Perturbed Einstein tensor for polar perturbation}
\label{app:TOV}
The components of perturbed field equations are
\begin{equation}\label{eq:55}
	\begin{split}
		\frac{\delta G^{t}_{t}}{Y^{0}_{l}\left(\theta,\phi\right)}
		=&\frac{2}{r^2}K+\frac{e^{-2\lambda}}{r}H'_{2}+e^{-2\lambda}\frac{r\lambda'-\left(l-1\right)\left(l+2\right)}{r}K'\\
		&+\left[\frac{\left(l-1\right)\left(l+2\right)+e^{-2\lambda}-2e^{-2\lambda}r\lambda'}{r^2}\right]H_{2}-e^{-2\lambda}K''
	\end{split}
\end{equation}
\begin{equation}\label{eq:56}
	\begin{split}
		\frac{\delta G^{r}_{r}}{Y^{0}_{l}\left(\theta,\phi\right)}
		&=-\frac{2}{r^2}K-\frac{e^{-2\lambda}}{r}H'_{2}+e^{-2\lambda}\frac{\left(l-1\right)\left(l+2\right)-r\psi'}{r}K'\\
		&+\left[\frac{\left(l-1\right)\left(l+2\right)-e^{-2\lambda}+2e^{-2\lambda}r\psi'}{r^2}\right]H_{0}+e^{-2\lambda}K''
	\end{split}
\end{equation}
\begin{equation}\label{eq:57}
	\begin{split}
	\frac{\delta G^{\theta}_{\theta}}{Y^{0}_{l}\left(\theta,\phi\right)}
	=&\frac{1}{2}e^{-2\lambda}H''_{2}-\frac{1}{2}e^{-2\lambda}K''
	+\frac{e^{-2\lambda}}{2}\left(\psi'-\lambda'-\frac{1}{r}\right)H'_{2}\\
	&-\frac{e^{-2\lambda}}{2}\left(\psi'-\lambda'+\frac{1}{r}\right)K'+\frac{1}{2r^2}\left(1-e^{-2\lambda}-2re^{-2\lambda}\psi'\right)H_{2}\\
	&-\frac{1}{2r^2}\left[\left(l-1\right)\left(l+2\right)-e^{-2\lambda}+2re^{-2\lambda}\psi'\right]K
	\end{split}
\end{equation}
\begin{equation}\label{eq:58}
	\begin{split}
	\frac{\delta G^{\phi}_{\phi}}{Y^{0}_{l}\left(\theta,\phi\right)}
	=&\frac{1}{r^2}\left[\left(l-1\right)\left(l+2\right)K-rK'\right]
	+\frac{1}{2}e^{-2\lambda}\left[H''_{2}+\left(\psi'-\lambda'+\frac{2}{r}\right)\right]H_2\\
	&-e^{-2\lambda}K''-e^{-2\lambda}\left[\psi'-\lambda'-\frac{1}{r}\right]K'+\frac{1}{2r^2}e^{-2\lambda}\left(1-2r\lambda'\right)H_{2}\\
	&-\frac{\left(l-1\right)\left(l+2\right)}{2r^2}H_{2}-\frac{1}{2r^2}
	\left[2+r\left(\psi'-\lambda'\right)\right]H_{0}
	\end{split}
\end{equation}
\begin{equation}\label{eq:59}
	\begin{split}
	\frac{\delta G^{r}_{\theta}}{\partial_\theta Y^{0}_{l}\left(\theta,\phi\right)}
	=&\frac{e^{-2\lambda}}{2}\left(\left[H'_{2}+H_{2}(\psi'+\lambda')-\frac{H'_{0}}{r}-\frac{K'}{r}\right]+\frac{e^{2\lambda}}{r^2}\left[H_{0}+H_{2}+2K\right]\right).		
	\end{split}
\end{equation}
Putting $l=2$ into equations \ref{eq:55} to \ref{eq:59} gives
\begin{equation}\label{eq:60}
	\begin{split}
		\frac{\delta G^{t}_{t}}{Y^{0}_{2}\left(\theta,\phi\right)}=&\frac{2K}{r^2}+\frac{e^{-2\lambda}}{r}H'_{2}+e^{-2\lambda}\frac{\left(-3+\lambda'r\right)}{r}K'+\frac{3+e^{-2\lambda}-2e^{-\lambda}r\lambda'}{r^2}H_{2}-e^{-2\lambda}K''
	\end{split}
\end{equation}
\begin{equation}\label{eq:61}
	\begin{split}
		\frac{\delta G^{r}_{r}}{Y^{0}_{2}\left(\theta,\phi\right)}=&-\frac{3}{r^2}H_{0}+\frac{2}{r^2}K+\frac{e^{-2\lambda}}{r}H'_{2}+e^{-2\lambda}\frac{\left(1+r\psi'\right)}{r^2}H_{0}+e^{-2\lambda}\frac{\left(-1-r\psi'\right)}{r}K'
	\end{split}
\end{equation}
\begin{equation}\label{eq:62}
	\begin{split}
		\frac{\delta G^{\theta}_{\theta}+\delta G^{\phi}_{\phi}}{Y^{0}_{2}\left(\theta,\phi\right)}=&e^{-2\lambda}\frac{\left[r\lambda'-r\psi'-2\right]}{2r}K'-\frac{1}{2}e^{-2\lambda}K''-\frac{3}{2r^2}H_{0}-\frac{e^{-2\lambda}\left[r\lambda'-2r\psi'-1\right]}{2r}H_{0}\\&+\frac{1}{2}e^{-2\lambda}H''_{0}+\frac{2re^{-2\lambda}\left[(-r\psi'-1)(\lambda'-\psi')+r\psi''\right]+3}{2r^2}H_{2}\\&-\frac{e^{-2\lambda}\left(-r\psi'-1\right)}{2r}H''_{2}
	\end{split}
\end{equation}		
\begin{equation}\label{eq:63}
	\begin{split}
		\frac{\delta G^{r}_{\theta}}{\partial_{\theta}Y^{0}_{2}\left(\theta,\phi\right)}=&-\frac{1}{2}e^{-2\lambda}H'_{0}+\frac{1}{2}e^{-2\lambda}K'+\frac{e^{-2\lambda}\left(-1-r\psi'\right)}{2r}H_{2}+\frac{e^{-2\lambda}\left(1-r\psi'\right)}{2r}H_{0}.
	\end{split}
\end{equation}

%%===================================================%%
%% For presentation purpose, we have included        %%
%% \bigskip command. Please ignore this.             %%
%%===================================================%%

\begin{appendices}

%%=============================================%%
%% For submissions to Nature Portfolio Journals %%
%% please use the heading ``Extended Data''.   %%
%%=============================================%%

%%=============================================================%%
%% Sample for another appendix section			       %%
%%=============================================================%%

%% \section{Example of another appendix section}\label{secA2}%
%% Appendices may be used for helpful, supporting or essential material that would otherwise 
%% clutter, break up or be distracting to the text. Appendices can consist of sections, figures, 
%% tables and equations etc.

\end{appendices}

%%===========================================================================================%%
%% If you are submitting to one of the Nature Portfolio journals, using the eJP submission   %%
%% system, please include the references within the manuscript file itself. You may do this  %%
%% by copying the reference list from your .bbl file, paste it into the main manuscript .tex %%
%% file, and delete the associated \verb+\bibliography+ commands.                            %%
%%===========================================================================================%%

\bibliography{sn-bibliography}% common bib file
%% if required, the content of .bbl file can be included here once bbl is generated
%\input sn-article.bbl

\end{document}